\documentclass[aps,prl,twocolumn,superscriptaddress]{revtex4}
\usepackage{graphicx}
\usepackage{amsmath,amssymb,latexsym}
\usepackage{bm}
\usepackage{epsfig}

\newcommand{\postscript}[2]{\setlength{\epsfxsize}{#2\hsize}
   \centerline{\epsfbox{#1}}}

\newcommand{\eps}{\epsilon}
\newcommand{\rar}{\rightarrow}
\newcommand{\barr}{\begin{array}}
\newcommand{\earr}{\end{array}}
\newcommand{\bea}[1]{\begin{eqnarray} \label{(#1)}}
\newcommand{\eea}{\end{eqnarray}}
\newcommand{\beq}[1]{\begin{equation} \label{(#1)}}
\newcommand{\eeq}{\end{equation}}

\newcommand{\rf}[1]{(\ref{(#1)})}

\newcommand{\rarr}{\longrightarrow}
\newcommand{\gr}{$\gamma$-ray}
\newcommand{\ngdd}{$A^*$}
\newcommand{\En}{E_N}

\newcommand{\sgdr} {\sigma^{\rm GDR}}
\newcommand{\ggdr} {\Gamma^{\rm GDR}}
\newcommand{\egdr} {\eps_\gamma^{\rm GDR}}
\newcommand{\edeex}{\eps_\gamma^{\rm dxn}}
\newcommand{\elab} {\eps_\gamma^{\rm LAB}}
\newcommand{\eamb} {\eps}
\newcommand{\ealab}{E_A^{\rm LAB}}

\newcommand{\Fcrab}{F_{\rm Crab}}
\newcommand{\xe}{x_\epsilon}

\newcommand{\Rsb}{R_{\rm SB}}

\begin{document}

\title{TeV ${\gamma}$ rays from photodisintegration and daughter
    deexcitation of cosmic-ray nuclei}

\author{Luis A.~Anchordoqui}

\affiliation{Department of Physics, University of Wisconsin-Milwaukee,
P.O. Box 413, Milwaukee, WI 53201
}

\author{John F. Beacom}
\affiliation{CCAPP, Departments of Physics and Astronomy,
Ohio State University,
Columbus, OH 43210
}

\author{Haim Goldberg}
\affiliation{Department of Physics,
Northeastern University, Boston, MA 02115
}

\author{Sergio Palomares-Ruiz}
\affiliation{Department of Physics and Astronomy,
Vanderbilt University, Nashville, TN 37235}
\affiliation{Institute for Particle Physics Phenomenology,
University of Durham, Durham DH1 3LE, UK
}

\author{Thomas J. Weiler}
\affiliation{Department of Physics and Astronomy,
Vanderbilt University, Nashville, TN 37235}

\begin{abstract}
  \noindent It is commonly assumed that high-energy $\gamma$-rays are
  made via either purely electromagnetic processes or the hadronic
  process of pion production, followed by decay.  We investigate
  astrophysical contexts where a third process (A*) would dominate,
  namely the photo-disintegration of highly boosted nuclei followed by
  daughter de-excitation.  Starbust regions such as Cygnus OB2 appear
  to be promising sites for TeV $\gamma$-ray emission via this
  mechanism.  A unique feature of the A* process is a sharp flattening
  of the energy spectrum below
  $\sim 10\,{\rm TeV/(T/eV)}$ for $\gamma$-ray emission from a
  thermal region of temperature~T.  We also check that a diffuse
  $\gamma$-ray component resulting from the interaction of a possible
  extreme-energy cosmic-ray nuclei with background radiation is well
  below the observed EGRET data. The A* mechanism described herein offers an important  
  contribution to $\gamma$-ray astronomy in the era of intense observational activity.
\end{abstract}


\maketitle


{\bf Introduction}.
In the field of TeV~\gr-astronomy, new instruments are discovering new
sources at a rapid rate, both within our Galaxy and outside the
Galaxy~\cite{TeVreview}.  Not surprisingly, one of the brightest TeV~\gr\
sources is the one discovered first, the Crab pulsar
wind nebula.  The integral \gr\ flux obtained from the Crab by the
Whipple Collaboration is now the standard TeV~flux unit:
$\Fcrab(\elab>0.35\,{\rm TeV})=10^{-10}{\rm /cm^2/s}$~\cite{crab}.
The spectral index of the Crab's integrated flux is measured to be
$-1.5$, so $\Fcrab$ falls by a factor of 30 for each decade of
increase in $\min (\elab)$. Newly commissioned atmospheric Cerenkov
telescopes (CANGAROO, HESS, MAGIC and VERITAS) will reach a
sensitivity 
100 times below $\Fcrab$.

Two well-known mechanisms for generating TeV $\gamma$-rays in
astrophysical sources~\cite{Aharonian:2004yt} are the purely
electromagnetic (EM) synchrotron emission and inverse Compton
scattering, and the hadronic (PION) one in which $\gamma$-rays
originate from $\pi^0$ production and decay.  There has been
considerable debate over which of these two mechanisms is dominant.
The mechanism for the PION mode may be either $pp$ or $p\gamma$
collisions.  In this Letter, we highlight a third dynamic which
leads to TeV $\gamma$-rays: photo-disintegration of high-energy nuclei,
followed by immediate photo-emission from the excited daughter nuclei.  For
brevity, we label the photo-nuclear process $A+\gamma\rar A'^*+X$,
followed by $A'^*\rarr A' +\gamma$-ray as ``$A^*$''.
It is likely that each of the three mechanisms is operative in some environments.
As we show below,
the $A^*$-process is likely operative in the interesting complex environments which have
accelerated nuclei and hot starlight.
Fortunately, future measurements can readily
distinguish among the three competing mechanisms per source.
In particular, the nuclei of the $A^*$ process act in analogy to a
relativistically moving mirror~\cite{movingmirror},
to ``double-boost'' eV~starlight to TeV energies for a nuclear boost factor
$\Gamma_A=\ealab/m_A > 10^6$.

{\bf The $\bm{A^*}$-Process}.
Nuclear photo-disintegration in the astrophysical context was first
calculated by Stecker in a seminal paper~\cite{Stecker:1969fw},
almost forty years ago.
To our knowledge, the possibly important role of
nuclear de-excitation in the astrophysical context
was first appreciated and proposed by Moskalenko and collaborators~\cite{BKM87} more
than a decade ago.
Since then, the \ngdd-process has been overlooked by the $\gamma$-ray community.
Now that data is becoming available which can validate or invalidate the process,
it is timely to revive and further develop the \ngdd-process.
We do so, by providing calculational details,
and by identifying the astrophysical context where this process
might dominate over the EM and PION modes for production of $\gamma$-rays.
A detailed discussion on the main features of the EM and PION mechanisms
is presented in a longer accompanying paper~\cite{LongVS}.

By far the largest contribution to the photo-excitation cross-section
comes from the Giant Dipole Resonance (GDR) at $\egdr \sim 10~{\rm
  MeV} - 30~{\rm MeV}$ in the nuclear rest frame~\cite{note2}.
The ambient photon energy required to excite the GDR is therefore
$\eamb = \egdr/\Gamma_A$.
The GDR decays by the statistical
emission of a single nucleon, leaving an excited daughter nucleus
$(A-1)^*$.  The probability for emission of two (or more) nucleons is
smaller by an order of magnitude.  The excited daughter nuclei
typically de-excite by emitting one or more photons of energies
$\edeex\sim 1-5$~MeV in the nuclear rest frame.  The lab-frame energy
of the $\gamma$-ray is then $\elab=\Gamma_A\,\edeex$.

As just outlined, the boost in the nuclei energy from rest to
$\ealab=\Gamma_A\,A\,m_N$ plays two roles.  It promotes the thermal
energies of the ambient photons to the tens of MeV $\gamma$-rays capable
of exciting the GDR, and it promotes the de-excitation photons from a
few MeV to much higher energies, potentially detectable in \gr\
telescopes.  Eliminating $\Gamma_A$ above leads to the relation
$\eamb\,\elab\sim\egdr\,\edeex$.  The ``$\sim$'' indicates that the
$\egdr$ and $\edeex$ energies are not sharp, but rather distributed
over their resonant shapes (``Lorentzian'' or ``Breit-Wigner''). Each
of the two spectra have a width to mass ratio less than unity. It is
sufficient for our purposes to treat each spectrum in the
narrow-width-approximation~(NWA).  Well-defined values for $\egdr$ and
$\edeex$ then result.  We may {\em summarize up to this point by saying
  that the \ngdd\-process produces \gr's with energy}
$\elab=\egdr\,\edeex/\eamb\sim\ 20\,{\rm TeV}/({\rm T/eV})$
{\em if there exists an accelerated nuclear flux with boost}
$\Gamma_A\sim\egdr/\eamb\sim 7\times 10^6\,({\rm T/eV})$
{\em or equivalent energy} $\ealab\sim 7\,{\rm PeV}/({\rm T/eV})\,{\rm
  per\; nucleon}$~\cite{BKM87,knee}.

Here and throughout we assume a Bose-Einstein (BE)
distribution with temperature ${\rm T}$ for the ambient photons; in
the mean, $\langle\eamb\rangle\sim 3\,{\rm T}$.  
As an example, \gr's with energy
$\elab\sim$~20\,TeV are generated in this \ngdd-process if an ambient
photon temperature of an eV and a boost factor of $7\times 10^6$ are
present.  Thus, we have arrived at an astrophysical environment where
the \ngdd-process may dominate production of TeV~\gr's: the region
must contain far-UV photons (commonly defined as 1-20~eV) 
from the Lyman-$\alpha$ emission of young,
massive, hot stars such as O and B stars which have surface
temperatures of 
${\rm T}_\star\sim 40,000$K 
(3.4~eV and $\langle\eps\rangle\sim10$~eV) 
and 18,000K (1.5~eV and $\langle\eps\rangle\sim 5$~eV),
respectively; and the region must contain shocks, giant winds, or
other mechanisms which accelerate nuclei to energies in excess of a
PeV per nucleon.  Violent starburst regions, such as the one in the
direction of Cygnus, are splendid examples of regions which contain OB
stars, shocks and giant stellar winds.

With $\egdr$ and $\edeex$ fixed at their central values, the \gr\
spectrum $dn(\elab)/d\elab$ is given by a simple Jacobian times the BE
distribution with argument $\eamb$ set to $\egdr\,\edeex/\elab$, i.e.,
\beq{inverseBE}
\frac{dn(\elab)}{d\elab}\propto
(\elab)^{-4}\,\left[e^{(\egdr\,\edeex/\elab\,{\rm T})}-1\right]^{-1}\,.
\eeq
Three regions in $\elab$ emerge.  For $\elab < \egdr\,\edeex/{\rm T}$,
the \gr\  spectrum is exponentially suppressed~\cite{note0}.
For $\elab \gg \egdr\,\edeex/{\rm T}$, there is power-law suppression.
The \gr\ spectrum peaks near $\elab\sim \egdr\,\edeex/{\rm T}$.  In
particular, one notes that the suppression of the
high-energy Wien end of the thermal photon spectrum has led to a
similar {\em suppression of the photon spectrum below
  $\elab\sim 20\,{\rm TeV}/({\rm T}/{\rm eV})$ }.  
This {\em lower-energy suppression presents
  a robust prediction of the $A^*$-process}.
Moreover, it contrasts greatly with the PION and EM processes, and so
provides a unique signature. The \ngdd\-process predicts ``orphan''
sources, with suppression of associated GeV \gr's or MeV X-rays (although the
photo-dissociated neutrons may $\beta$-decay to
neutrinos~\cite{n2nu}). Observationally, orphan \gr\ sources have been
identified~\cite{HEGRA05,orphans}, and some orphans are known to be near
OB~starburst regions~\cite{HESSorphans}.

Comparing to the PION~$p\gamma$ process,
the \ngdd\ process has a much lower ambient photon energy threshold;
in the nuclear rest frame, the photon threshold is $\egdr\sim 10$~MeV for the
\ngdd\ process, but it is an order of magnitude larger at $m_\pi$
for the PION~$p\gamma$ process~\cite{Stecker:1998ib}.  This means that for fixed
ambient temperature, $\Gamma_A$ need be an order of magnitude larger for the
PION~$p\gamma$ process, and the resulting \gr\ energies, proportional to $\Gamma_A^2$,
are two orders of magnitude larger.  The EM and PION~$pp$ processes contrast
with the $A^*$-process in that there is either no threshold (EM)
or very small threshold ${\cal O} (2m_\pi)$ in the lab (PION~$pp$),
and the resulting \gr\ spectrum
rises monotonically with decreasing $\elab$.

{\bf The $\bm{A^*}$-Rate}.
We now derive the rate for \gr\ production in the \ngdd-process.
The cross-section is dominated by the GDR dipole
form~\cite{Stecker:1998ib}, which in the NWA is
\beq{sigmaGDR}
\sigma_A (\eamb)
\stackrel{\rm
  NWA}{\rarr}\,\frac{\pi}{2}\,\sgdr\,\ggdr\,
\delta(\egdr-\Gamma_A\,\eamb)\,,
\eeq
where $\ggdr$ and $\sgdr$ are the GDR width and cross-section at
maximum. Fitted numerical formulas are $\sgdr = 1.45\,A\times 10^{-27}
{\rm cm}^2$, $\ggdr = 8~{\rm  MeV}$, and  $\egdr = 42.65
A^{-0.21}$~MeV for $A > 4$ and $0.925 A^{2.433}$ for $A\leq
4$~\cite{KT93}.
The 8~MeV width implies a very short de-excitation distance of $\sim 25\,\Gamma_A$~fm.

The general formula for the inverse photo-disintegration
mean-free-path (mfp)~\cite{note3} for a highly relativistic nucleus
with energy $\Gamma_A/m_A$ propagating through an isotropic photon background
with energy
$\eps$ and spectrum $dn(\eps)/d\eps$
is~\cite{Stecker:1969fw,LongVS}
\beq{stecker}
(\lambda_A)^{-1}
   \stackrel{\rm NWA}{\rarr}\,
\frac{\pi \, \sgdr \,\egdr\, \ggdr}{4\,\Gamma_A^2}
\int_{\frac{\egdr}{2\,\Gamma_A}}^\infty
\frac{d\epsilon}{\epsilon^2}\,\frac{dn(\eps)}{d\eps}\,.
\eeq
For a nucleus passing through a region of thermal photons, 
integration over the BE photon distribution gives  
\beq{ImfpBE}
(\lambda_A^{\rm BE})^{-1} \approx \frac{\sgdr\,\ggdr}{\egdr}\,
n_\gamma^{\rm BE}
\,w^2\, |\ln \left(1 -
e^{-w}\right)| \,,
\eeq
where we have defined a dimensionless scaling variable $w\equiv
\egdr/2\,\Gamma_A\,{\rm T}$. 
From the prefactor, we learn that the peak of $(\lambda_A^{\rm BE})^{-1}$
scales in $A$ as $\sgdr/\egdr\sim A^{1.21}$, 
and the value of $\Gamma_A$ at the peak scales as $\egdr\sim A^{-0.21}$.

\begin{figure}[thb]
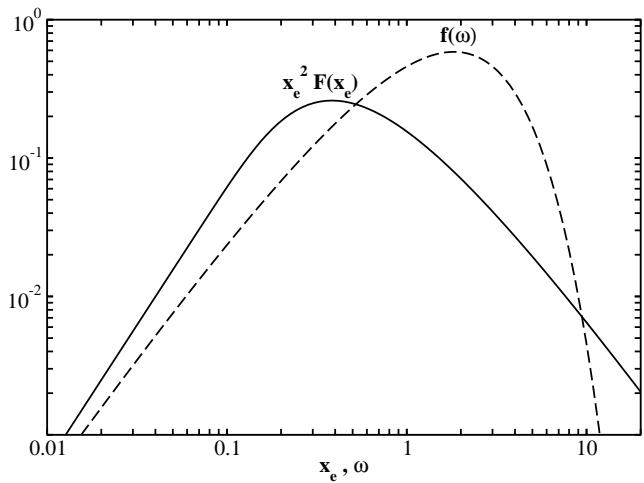

\begin{center}
\postscript{Fxfwlog2b.eps}{0.98}
\caption{\label{fig:one}
The scaling functions $f(w)$ 
and ${\xe}^2 F(\xe)$,
proportional to the photo-dissociation rate in Eq.~(4)
and the \gr\ $\eps^2\,dF/d\eps$ spectrum in Eq.~(8),
respectively.
}
\end{center}
\end{figure}

The scaling function 
$f(w)=w^2\, |\ln \left(1 - e^{-w}\right)|$ is shown in Fig.~\ref{fig:one}. 
Approximations to
the $|\ln|$~term yield $e^{-w}$ for $w>2$, and $|\ln w|$ for $w\ll
1$. Thus, the exponential suppression of the process appears again for
large for $w>2$, i.e., for $\elab < \egdr\,\edeex/4\,{\rm T}$, and the
small $w$ region presents a mfp that scales as $w^2\,|\ln w|$.
The peak region provides the smallest inverse mfp, and so this region
dominates the \ngdd-process.  In the peak region, $w$ is of order
one, which implies that $\Gamma_A\,{\rm T}\sim \egdr$. When this
latter relation between the nuclei boost and the ambient photon
temperature is met, then the photo-disintegration rate is optimized.

The area under the peak region in Fig.~\ref{fig:one} is of order one,
which leads to a simple and reasonable estimate of the inverse mfp
given in Eq.~\rf{ImfpBE}, for any nucleus boosted near $\Gamma_A
\sim\egdr/{\rm T}$. The useful and eminently sensible estimate is 
$(\lambda_A^{\rm BE})^{-1} \approx \sgdr\,N_\gamma^{\rm BE}$;
here, we have input the typical value $\ggdr/\egdr\sim 1/3$. 
Putting in numbers, this estimate yields
\beq{mfp}
\lambda_A^{\rm BE}\sim \frac{5\times 10^{13}\,{\rm cm}}{A\,\left({\rm
    T/eV}\right)^3} \sim \frac{3\,{\rm AU}}{A\,\left({\rm T/eV}\right)^3}\,
\eeq
for a nucleus with energy in the peak regime around $E_A\sim
10\,A/({\rm T/eV})$~PeV.
Here we have used 
$n_\gamma^{\rm BE}=2\,\zeta(3)\,{\rm T}^3/\pi^2\simeq {\rm T}^3/4$. 

An important question is how many photo-disintegration steps in the
nuclear chain $A\rightarrow (A-1)^* \rightarrow (A-2)^*$ etc.\ may be
expected.  Each step will produce an excited daughter which may
de-excite via emission of \gr's, each having a typical lab-frame energy
$\elab\sim\egdr\,\edeex/3\,{\rm T}$.  Clearly, the number of steps
depends on the mfp of each excited daughter nucleus, and on the
diffusion time of the nuclei in the thermal region.  The probability
for $n$-cascades, producing $n$ excited daughter nuclei and $n$ single
nucleons, may be written in symmetric ways as
\beq{cascade1}
P_n  =  \prod_{j=1}^n \int^{x_{j+1}}_0 \frac{dx_j}{\lambda_j}
e^{-\frac{(x_j-x_{j-1})}{\lambda_j}} = \prod_{j=1}^n \int_0^{x_{j+1}}
\frac{dx_j}{\lambda_j} e^{-x_j\,\delta_j} ,
\eeq
where $x_0=0$, $x_{n+1}$ equals the diffusion length $D$ of nuclei in
the $A^*$-region, the ordered $x_1 \le x_2 \ldots x_n $ denote the
spatial positions of successive photo-disintegrations to $(A-1)^*$,
$\ldots\,(A-n)^*$, $\delta_j=\lambda_j^{-1}-\lambda_{j+1}^{-1}$, and
$\lambda_{n+1}=\infty$. The exponentials in \rf{cascade1} are
probabilities that the various daughters not interact (i.e., survive)
from the point of creation to the point of photo-disintegration. A
simple result obtains when the mfp's are long on the diffusion scale
$D$ of the $A^*$-region.  In this case the nested integrals collapse to
$P_n\approx (n!)^{-1}\,\prod_{j=1}^n \,D/\lambda_j$. This is just the
product of the independent probabilities for each excited daughter to
be produced, times the factor $1/n!$ that divides out all but the one
correct time ordering of the $n$~photo-disintegrations.  One sees
that, since $D/\lambda_j \ll 1$ by assumption, disintegration to more
than the first excited daughter is unlikely.  Such is the case if
photo-disintegration of nuclei occurs in the Cygnus OB starburst
region, as we now demonstrate.

{\bf Relevance to Starburst Regions}.
The aforementioned possibility of generating Galactic TeV \gr's from
accelerated nuclei scattering on starlight in starburst regions is of
considerable astrophysical interest.  The energy of the photons ($\sim
3{\rm T}_\star$ in the mean) is maintained as the photons disperse from the stars,
but the photon density decreases by the ratio of the surface of
stellar emission ($N_{\star}\times 4\,\pi\,R_{\star}^2$) to the loss
surface of the starburst region ($4\,\pi\,\Rsb^2$).  Taking the giant
star radius to be $R_{\star}\sim 10\,R_\odot$, the radius of the
starburst region to be $\Rsb\sim 10$~pc, and the stellar count to be
$N_{\star}\sim 2600$, the photon density is diluted by $\sim 10^{12}$.
According to Eqs.~\rf{stecker} and \rf{ImfpBE}, 
$(\lambda_A)^{-1}$ is proportional to
the photon density, and hence to this factor.  Including this factor
in Eq.~\rf{mfp}, one gets the estimate
$\lambda_A\sim (56/A)\,(1.5\,{\rm eV/T}_\star)^3 \times 10^{23}$~cm
for the nuclear mfp in the starburst region.  
The hot stars are dominantly (95\%) B type, with
${\rm T}_\star\sim 1.5$~eV.  We note here that the diffusion time for a nucleus in the
Cygnus OB starburst region is calculated to be $\sim 10^4$~yr $\sim
10^{22}$~cm, an order of magnitude shorter than the nuclear mfp.
Accordingly, one expects only a few percent of the nuclei to
photo-disintegrate in the Cygnus OB region, with multiple
disintegrations being rather rare.

Some early statistical-model calculations for the production of
$\gamma$-rays through the decay of the GDR gave a mean photon
multiplicity between 0.5 and 2 for the different nuclides in the
chain reaction (see Ref.~\cite{LongVS} for details).  We will simplify the
data by assuming that one photon is emitted per nuclear
de-excitation. Then, the rate of photo-emission is just the rate of
excited daughter production, which in turn is just the rate of
photo-disintegration.  Allowing for the $1/r^2$ dilution of the \gr\
flux from the source region of volume $V_{A^*}$ at distance $d$, the
observed integral \gr\ flux ${\cal F}_\gamma$ at Earth with $\elab$
above a few TeV $\times$~(eV/T) is then
\beq{fluxEarth}
{\cal F}_\gamma = \frac{V_{A^*}}{4\pi\,d^2}\,\frac{1}{\lambda^{\rm
    BE}_A}\,{\cal F}_A\,,
\eeq
where the flux ${\cal F}_A (=c\,n_A)$ is integral over the
energy-decade of the peak region,
$A\,{\rm(eV/T)\,PeV}<E_A< 10\,A\,{\rm (eV/T)}$.
It is commonly assumed that this nuclear flux results
from continuous trapping of the diffuse cosmic ray flux by diffusion
in a milligauss magnetic field.

Putting into Eqs.~\rf{mfp} and \rf{fluxEarth} the parameters for the Cygnus OB2 region,
i.e., $\lambda_A=(56/A)\times 10^{23}$~cm, $\Rsb=10$~pc, and $d=1.7$~kpc, 
one obtains ${\cal F}_\gamma = 2\times 10^{-10}\,A\,{\cal F}_A$.  Thus, an
accumulated nuclear flux within Cygnus OB2 of $f{\rm /cm^2/s}$ in the
PeV region gives a $\ge$~TeV \gr\ flux at Earth of $\sim 10\,A\,f\times
\Fcrab (\elab\ge 1\,{\rm TeV})$.  For example, an iron flux
${\cal  F}_{56}=6\times 10^{-5}{\rm /cm^2/s}$ above a TeV leads to a \gr\ flux at Earth of about
3\% of the Crab, just at the level measured by the HEGRA experiment
for \gr's from the Cygnus OB2 direction~\cite{HEGRA05}.
In our accompanying paper~\cite{LongVS} we present a more detailed calculation of
the $A^*$-process for this starburst region,
and show that ${\cal F}_{56}\sim 10^{-4}{\rm /cm^2/s}$
is a credible 1\% of the kinetic energy budget of Cygnus~OB2.

The \gr\ energy spectrum remains to be discussed. In the rest frame
of the excited nucleus, the photon is emitted isotropically and nearly 
monochromatically. Therefore, in the lab frame, 
the \gr-spectrum is nearly flat between 0 and $2\,\Gamma_A\,\edeex$ on a linear scale.
%
%
The power spectrum and integrated power
rise as $\elab$ and $({\elab})^2$, respectively,
peaking near 
$\elab \sim \egdr\,\edeex/4\,{\rm T}\sim 10\,{\rm TeV}\,({\rm eV/T})$,
as explained in the paragraph below Eq.~\rf{ImfpBE}.

In detail, the photon spectrum is obtained by replacing the approximate Eq.~\rf{fluxEarth} 
with an integral over $(dF_A/d\ealab) \times (\lambda_A^{\rm BE})^{-1}$,
with measure
$d\ealab\;d\cos\theta_\gamma\;d\elab\;
   \delta [\elab-\Gamma_A\,\edeex\,(1+\cos\theta_\gamma)]$,
where $\theta_\gamma$ is the angle in the nucleus rest frame 
between the isotropically-emitted photon and the boost direction.
After assuming a power-law (with spectral index $\alpha$) for the nuclear spectrum,
and integrating over $d\cos\theta_\gamma$, one arrives at
\bea{photon-spectrum}
\frac{(\elab)^2\,d{F}_\gamma (\elab)}{d\elab} =
\frac{n^{\rm Th}\,V_{A^*}}{4\pi\,d^2}\,
\frac{\sgdr\,\ggdr\,\edeex\,m_N}{\egdr}\, \\
\times 
    \left(\frac{\rm T}{\egdr}\right)^{\alpha-2}\,
    \left[\left(\frac{2\En}{m_N}\right)^\alpha 
    \frac{dF_N}{d\En}\right]_{E_0}\,{\xe}^2\,F(\xe)\,, \nonumber
\eea
where $n^{\rm Th}$ is the true density of the thermal photons after spatial dilution, 
$E_0$ is any reference energy,
$\xe=\elab\;{\rm T}/ \edeex\,\egdr$ is the dimensionless 
energy-scaling variable,
and 
\beq{scaling-function}
F(\xe)=
     \int \frac{d\omega}{\omega}\,\omega^\alpha\,f(\omega)
   = \int_0^{1/{\xe}} d\omega\,\omega^{1+\alpha}\,|\ln (1-e^{-w})|
\eeq
is the scaling function.
Shown in Fig.~(\ref{fig:one}) is ${\xe}^2\,F(\xe)$~\cite{noteN}. 
The \gr\ energy is $\elab= 10\,\xe\,{\rm TeV}\times 
(\edeex/{\rm MeV})\,(\egdr/10\,{\rm MeV})\,({\rm T/eV})^{-1}$.
We note that the prefactor of ${\xe}^2\,F(\xe)$ 
in Eq.~\rf{photon-spectrum} scales in $A$ for fixed energy per nucleon 
as $\sgdr\times (\egdr)^{-(\alpha-1)}\propto A^{1+0.21(\alpha-1)}$,
while the position of the peak in the ${\xe}^2\,F(\xe)$ spectrum 
at $\xe\sim 0.25$ scales mildly as $\xe\propto\egdr\propto A^{-0.21}$.
Well beyond the peak, 
the ${\xe}^2\,F(\xe)$ spectrum falls as 
$\xe^{-2}\,\ln\xe$; equivalently, the \gr\ spectrum falls 
as $\approx (\elab)^{-4}$.

{\bf The Cosmogenic $\bm{A^*}$-Process}.
One well-known application~\cite{Stecker:1998ib} of nuclear
photo-disintegration occurs in the calculation of the propagation of
extreme-energy cosmic nuclei in the cosmic microwave background
(CMB) and cosmic infrared background (CIRB).  The CMB contribution is
thermal, with temperature ${\rm T} = 2.3 \times 10^{-4}$~eV, and
Eq.~\rf{mfp} readily yields a mfp of $\sim A^{-1}$~Mpc for nuclei with
energies near $E_A\sim 4\,A\times 10^{19}$~eV.  The CIRB, now fairly
well known~\cite{CIRB}, is much smaller than the CMB. Its spectrum
may be approximated by one or two power-laws, which allows an analytic
integration of Eq.~\rf{stecker}.  The result is a mfp longer than that
of the CMB result, but relevant to nuclei with energies down to
$10^{16}$~eV.

Of interest in our work is the \gr\ flux produced when the
photo-dissociated nuclear fragments produced on the CMB and CIRB
de-excite.  These $\gamma$-rays create chains of electromagnetic cascades on
the CMB and CIRB, resulting in a transfer of the initial energy into
the so-called EGRET region below 100~GeV, which is bounded by
observation to not exceed $\omega_{\rm cas}\sim 2\times 10^{-6} {\rm
  eV/cm^3}$~\cite{Sreekumar:1997un}.

Fortunately, we can finesse the details  of the calculation by arguing
in analogy to work already done.  Two recent papers~\cite{HTS,ABOWY}
have calculated the ${\bar\nu}_e$ flux resulting from
photo-disintegration of cosmic Fe, followed by $\beta$-decay of the
associated free neutrons.  The photo-disintegration chain produces one
$\beta$-decay neutrino with energy of order 0.5~MeV in the nuclear rest
frame, for each neutron produced. Multiplying this result by 2 to
include photo-disintegration to protons in addition to neutrons
correctly weights the number of steps in the chain.  Each step produces on average
one photon with energy $\sim 3$~MeV in the nuclear rest frame.
Comparing, about 12 times more energy is deposited into
photons. Including the factor of 12 relating $\omega_\gamma$ to
$\omega_\nu$, we find from Fig.~3 in Ref.~\cite{HTS} that cosmogenic
photo-disintegration/de-excitation energy is more than three orders of magnitude
below the EGRET bound~\cite{DHooper}.  This result appears to be
nearly invariant with respect to varying the maximum energy of the Fe
injection spectrum (with a larger $E_{\rm max}$, the additional energy
goes into cosmogenic pion production).  Thus, there is no constraint on
a heavy nuclei cosmic-ray flux from the $A^*$ mechanism.

{\bf Conclusion}.
In final summary, we have presented an alternative mechanism ($A^*$)
for generating TeV \gr's in starburst regions.  It has a unique
orphan-like signature, with a flat spectrum below 
$\sim 20\,{\rm TeV/(T/eV)}$ and a quasi power-law above. 
Since the EM and PION mechanisms also produce unique signatures,
gamma-ray astrophysics benefits from a one-to-one correspondence between the 
source dynamic and the observable spectrum.

  We acknowledge encouragement from V. Berezinsky, 
  useful comments from R. Ong,
  and grant support from NSF PHY-0547102 (CAREER, JFB), 
  PHY-0244507 (HG), 
  NASA ATP02-000-0151 (SPR and TJW), 
  Spanish MCT FPA2005-01678 (SPR), 
  DOE DE-FG05-85ER40226 (TJW) and Vanderbilt University Discovery Award (TJW).


\begin{thebibliography}{99}
\bibitem{note2}
  Note that we will use $\eps$ for photon energies, and $E$ for hadron
  and neutrino energies.  It is also implicit that GDR-superscripted
  variables ($\egdr$ here, and $\ggdr$ and $\sgdr$ below) have an
  $A$-dependence.
\bibitem{note0} 
This suppression characterises a mean energy.
However, as explained later in the text, at low energies the differential spectrum
is actually flat, so 
the integrated power at low-energy is suppressed as $(\elab)^2$.
\bibitem{noteN}
  The approximate reflection symmetry in Fig.~(\ref{fig:one}) between
$\xe^2\,F(\xe)$ and $f(w)$ about the value $\xe=w=1/2$ is due in part to the 
inverse relation $\xe\,w = \elab/2\,\Gamma_A\,\edeex = (1+\cos\theta_\gamma)/2$,
which equals $1/2$ in the mean.
\bibitem{note3}
  $\lambda_A=\tau_A$ is also the mean survival time, so
  $\lambda_A^{-1}$ is also the photodisintegration rate per nucleus.
\end{thebibliography}
\end{document}